\documentstyle[12pt,epsf,fleqn]{elsart}

\def \ni{\noindent}

\def \be {\begin{equation}}
\def \ee {\end{equation}}

\begin{document}

\ni {\bf Corresponding author}\\
Prof. Dr. H.V. Klapdor-Kleingrothaus\\
Max-Planck-Institut f\"ur Kernphysik\\
Saupfercheckweg 1\\
D-69117 HEIDELBERG\\
GERMANY\\
Phone Office: +49-(0)6221-516-262\\
Fax: +49-(0)6221-516-540\\
email: klapdor@gustav.mpi-hd.mpg.de\\

\begin{frontmatter}
\title{'Naked' Crystals go Underground}

\author{H.V. Klapdor-Kleingrothaus}
\footnote{
 Home-page: $http://www.mpi-hd.mpg.de.non\_acc/$}

\address{Max-Planck-Institut f\"ur Kernphysik, PO 10 39 80,
  D-69029 Heidelberg, Germany}

\date{07.05.2003}
\end{frontmatter}

\begin{figure}[ht]

\epsfysize=75mm\centerline{\epsffile{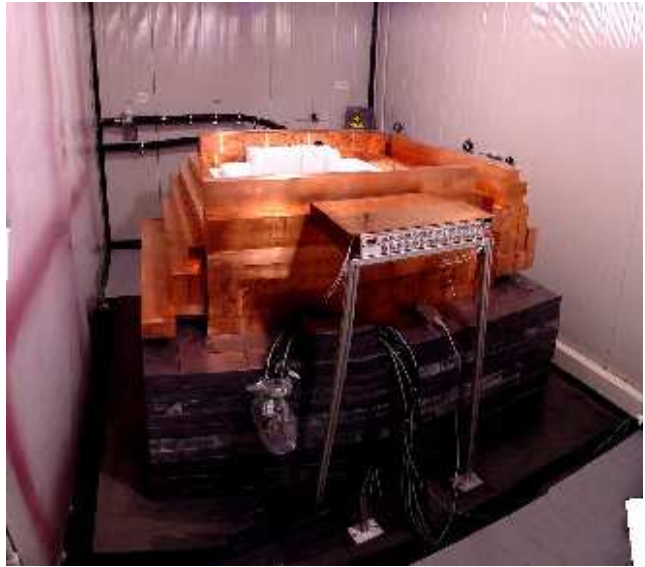}}
\caption[]{
	View of GENIUS-TF in the Gran Sasso Underground Laboratory in Italy.
	}
\label{fig:GENIUS-TF}
\end{figure}

\begin{figure}[ht]
\epsfysize=75mm\centerline{\epsffile{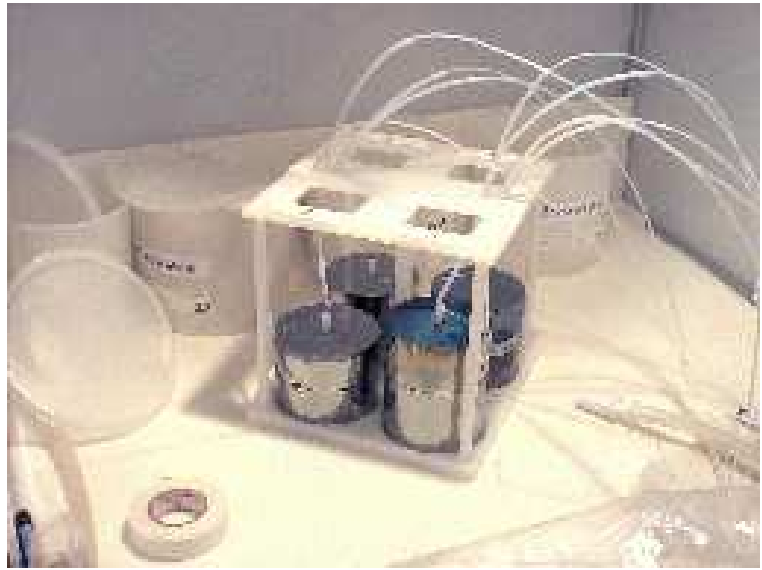}}
\caption[]{
	The contacted four naked detectors in the low-level holder 
	in which they are put into the shielded liquid nitrogen 
	container of GENIUS-TF.
	}
\label{fig:4Nak-Det}
\end{figure}

	On May 5, 2003 in the GRAN SASSO Underground Laboratory 
	the first naked high-purity Germanium detectors were 
	installed successfully in liquid
	nitrogen in the GENIUS-Test-Facility (GENIUS-TF) (Figs. 
\ref{fig:GENIUS-TF}, \ref{fig:4Nak-Det}). 
	This is the first time ever that this novel technique 
	for extreme background reduction in
	searches for rare decays  is going to be tested under 
	realistic background conditions.\\

\begin{figure}[ht]

\epsfysize=55mm\centerline{\epsffile{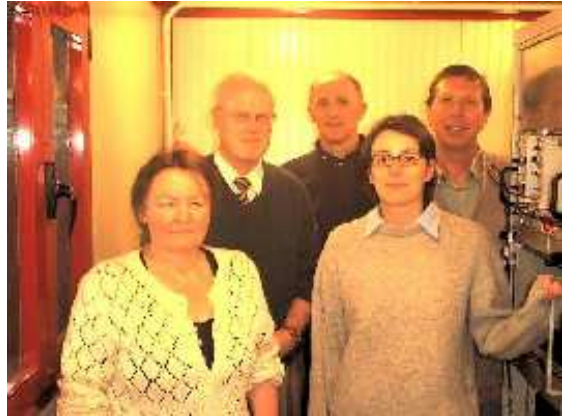}}
\caption[]{
	The successful team after installation of the detectors, 
	on May 5, 2003. 
	From left to right: Irina Krivosheina, 
	Hans Volker Klapdor-Kleingrothaus
	(Spokesman of the Heidelberg-Moscow experiment and also speaker of
	this collaboration), Oleg Chkvorets, Irina Krivosheina,  
	Claudia Tomei and Herbert Strecker from the Max Planck Institute 
	for Nuclear Physics in Heidelberg.
	}
\label{fig:team-TF}
\end{figure}

	The team which achieved this consists 
	of Hans Volker Klapdor-Kleingrothaus
	(Spokesman of the Heidelberg-Moscow experiment and also speaker of
	this collaboration), Oleg Chkvorets, Irina Krivosheina, 
	Herbert Strecker and Claudia Tomei from the Max Planck Institute 
	for Nuclear Physics in Heidelberg (
Fig. 
\ref{fig:team-TF}).\\

	The naked crystals are sitting on a plate made from a special type 
	of teflon, in a thin-walled copper
	box filled with ~70 liters of highly purified nitrogen.
	The copper is thermally shielded by 20 cm of special 
	low-level styropor, followed by a shield of 15 tons 
	of electrolytic copper (10 cm), and 35 tons of lead (20 cm).\\

	The total setup is being shielded by 10 cm of Boronpolyethylene as a
	neutron shield. A new digital data acquisition system allows to
	simultaneously measure energy, pulse shapes, and other 
	parameters of the individual events.

\begin{figure}[ht]

\epsfysize=75mm\centerline{\epsffile{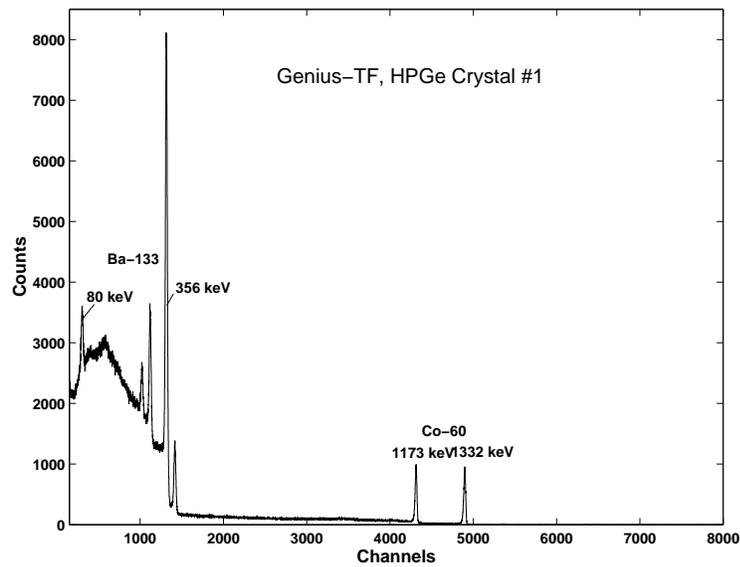}}
\caption[]{
	Spectrum measured for one of the detectors with a $^{60}{Co}$ 
	and a $^{133}{Ba}$ source. 
	The other detectors show the same quality.
	}
\label{fig:Ba}
\end{figure}

	The four detectors, in total 10 kg of high-purity natural Germanium,
	have been tested in the day of installation already, with radioactive
	sources of $^{60}{Co}$ and  $^{228}{Th}$ and show good energy 
	resolution. 
	A spectrum is shown in 
Fig. 
\ref{fig:Ba}. 
	It can be said already also that microphonics in the liquid
	nitrogen is not a problem.

	The GENIUS-TF project of the HEIDELBERG group
	had been approved in Gran Sasso in early 2001 -
	after the first Germanium crystals used in liquid nitrogen for
	spectroscopy had been tested already in Heidelberg
	in 1997 (J. Hellmig and H.V. Klapdor-Kleingrothaus, 
	Z. Phys.A 359 (1997) 351, 
	H.V. Klapdor-Kleingrothaus, J. Hellmig and M. Hirsch, J. 
	of Physics G, 24 (1998) 483, in connection with the proposal 
	of GENIUS (CERN COURIER Dec. 1997, 
	H.V. Klapdor-Kleingrothaus, Int. J. of Physics A 13 (1998) 3953).

	GENIUS was proposed to look with extreme sensitivity for cold dark
	matter (CDM), double beta decay and low energy solar neutrinos.

	With the successful start of operation of Ge detectors 
	in liquid nitrogen in Gran Sasso a historical step has been 
	achieved of a novel technique,
	and into a new domain of background reduction in underground physics
	in the search for rare events. Besides testing of constructional 
	parameters for the GENIUS project, 
	one of the first goals of GENIUS-TF will be to test the signal 
	of cold dark matter reported by the DAMA collaboration 
	a few years ago (Bernabei et al.) 
	which could originate from modulation of the WIMP flux 
	by the motion of the Earth relative to that of the Sun.

	Hans Volker Klapdor-Kleingrothaus


\end{document}